\documentclass[a4paper]{article}
\def\be{\begin{equation}}
\def\eea{\end{eqnarray}}
\def\bea{\begin{eqnarray}}
\def\ee{\end{equation}}
\def\a{\alpha}
\def\b{\beta}
\def\g{\gamma}
\def\s{\sigma}
\usepackage[psamsfonts]{amssymb}
\usepackage{amsmath}
\author{Masoud Alimohammadi \footnote{alimohmd@ut.ac.ir}
\\ {\small Department of Physics, University of Tehran,}
\\ {\small North Karegar Ave., Tehran, Iran.}}
\title{Solvable multi-species reaction-diffusion processes,including the extended drop-push model}
\begin{document}
\maketitle
\begin{abstract}
\noindent By considering the master equation of asymmetric
exclusion process on a one-dimensional lattice, we obtain the most
general boundary condition of the multi-species exclusion
processes in which the number of particles is constant in time.
This boundary condition introduces the various interactions to the
particles, including ones which have been studied yet and the new
ones. In these new models, the particles have simultaneously
diffusion, the two-particle interactions $A_\alpha
A_\beta\rightarrow A_\gamma A_\delta$, and the $n$-particle
extended drop-push interaction. The constraints on reaction rates
are obtained and in two-species case, they are solved to obtain a
solvable model. The conditional probabilities of this model are
calculated.

\end{abstract}
\section{Introduction}
One-dimensional asymmetric simple exclusion processes (ASEP) have
been shown to be of physical interest in various problems in
recent years. These problems are, for example: the kinetics of
biopolymerization \cite{1}\cite{n1}, traffic models \cite{2},
polymers in random media, dynamical models of interface growth
\cite{3}\cite{n2}, noisy Burgers equation \cite{4}, and study of
the shocks \cite{5}\cite{6}. There are many review articles in
these fields, see for example \cite{n3}-\cite{ch}.

The totally ASEP has been solved exactly in ref.\cite{7}. In this
simple model, each lattice site is occupied by at most one
particle and particles hop with rate one to their
right-neighboring sites if they are not occupied. The model is
completely specified by a master equation and a boundary
condition, imposed on probabilities appear in the master equation.
The coordinate Bethe ansatz has been used to show the
factorization of the $N$-particle scattering matrix to the
two-particle matrices.

By choosing other suitable boundary conditions, without changing
the master equation, one may study the more complicated
reaction-diffusion processes, even with long-range interaction. In
ref.\cite{8}, the so-called drop-push model has been studied by
this method. In this model the particle hops to the next right
site even it is occupied. The particle hops to this site by
pushing all the neighboring particles to their next right sites,
with a rate depending on the number of right neighboring
particles. The generalization of this model, by considering both
the right and left hopping, has been done in ref.\cite{9}. Further
generalization of the ASEP boundary condition have been proposed
in \cite{s1}-\cite{am}. As a comprehensive review on the ASEP, see
\cite{sc}.

All of the above studies are about the single-species systems. If
one considers the model with more-than-one species, the situation
becomes more complicated. The source of complexity is the
abovementioned factorization of $N$-particle scattering matrix,
which in these cases restricts the two-particle $S$-matrices
satisfies in some kind of spectral Yang-Baxter (SYB) equation. In
ref.\cite{10}, all the solvable two-species reaction-diffusion
models, in which the number of particles is constant in time and
the reaction rates are all equal, have been obtained. It was shown
that there are 28 independent interactions (among 4096 possible
types) which are solvable. One of these 28 models, in which the
particles have exchange-interaction: $A+B\rightarrow B+A$, has
been generalized to $p$ species in ref.\cite{11}. In this model,
besides diffusion to the right (with equal rates), particles
interact through $A_j+A_i\rightarrow A_i+A_j$ with rate $r_{ij}$.
The spectrum of $r_{ij}$, to ensure the solvability, has been
obtained. Another multi-species reaction-diffusion model, which
somehow relates to one considered in \cite{11}, has been discussed
in \cite{n9}. In this model the particles have
exchange-interaction with rates which are determined by the
differences of their diffusion rates.

The multi-species generalization of the models considered in
\cite{10}, has been considered in \cite{12}. The processes are
 \bea\label{1}
 A_\alpha\emptyset &\rightarrow &\emptyset A_\alpha \ \ \ \ {\rm with
 \ rate}\ \ 1,\cr
  A_\alpha A_\beta &\rightarrow & A_\gamma A_\delta  \ \ \ \ {\rm with
  \ rate} \ \ c^{\alpha\beta}_{\gamma\delta}.
 \eea
Again the solvability restricts the reaction rates
$c^{\alpha\beta}_{\gamma\delta}$. Some general remarks on the
solution of SYB equation, and some special solutions of it have
been given in \cite{12}.

Recently, the generalization of the drop-push model to
multi-species case has been considered in ref.\cite{13}. The
reactions are
 \bea\label{2}
 A_\alpha\emptyset &\rightarrow &\emptyset A_\alpha \ \ \ \ {\rm with
 \ rate}\ \ 1,\cr
 A_\alpha A_\beta\emptyset &\rightarrow &\emptyset A_\gamma A_\delta  \ \ \ \ {\rm with
  \ rate} \ \ b^{\alpha\beta}_{\gamma\delta}.
 \eea
The same drop-push reactions, but with $n$ adjacent particles and
with rates that are specified by a specific combinations of
$b^{\alpha\beta}_{\gamma\delta}$'s, also exist. Some comments have
been given for solutions of SYB equation in \cite{13}. The
reactions (\ref{2}) are called the extended drop-push reactions.

In this paper we are going to study the most general multi-species
model, i.e. the most general boundary condition, which {\it all}
the previous mentioned models are the special cases of it. In its
general form, the reactions are
 \bea\label{3}
 A_\alpha\emptyset &\rightarrow &\emptyset A_\alpha \ \ \ \ {\rm with
 \ rate}\ \ 1,\cr
  A_\alpha A_\beta &\rightarrow &A_\gamma A_\delta  \ \ \ \ {\rm with
  \ rate} \ \ c^{\alpha\beta}_{\gamma\delta},\cr
  A_\alpha A_\beta\emptyset &\rightarrow &\emptyset A_\gamma A_\delta  \ \ \ \ {\rm with
  \ rate} \ \ b^{\alpha\beta}_{\gamma\delta},\cr
   &\vdots &
    \eea
where the dots indicates the other drop-push reactions with $n$
adjacent particles, in which in the meantime the types of the
particles can also be changed. We show that the reaction rates
must satisfy some specific constraints, in order that we have a
set of consistent evolution equations. The two-particle
$S$-matrices of this general multi-species model must also satisfy
SYB equation, which in this case becomes more complicated. In the
two-species case, we study a special class of reactions (\ref{3})
in detail and show that in this class, there is only {\it one}
solvable model, i.e. the solution of SYB equation is unique in
this case. We study some physical properties of this unique model.

The scheme of the paper is the following. In section 2, by using
the law of conservation of probabilities, we obtain the most
general boundary condition for a $p$-species reaction-diffusion
model, in terms of two $p^2\times p^2$ matrices $b$ and $c$. We
obtain a constraint on the sum of the elements of each column of
matrix $b+c$. We see that in special cases, this model is same as
those have been studied previously. In section 3, we consider the
matrix $c$ as a diagonal matrix. It is shown that there are two
cases. In the first case, the matrix $b$ must be also diagonal,
which the model reduces to the ordinary, i.e. single-species,
drop-push model with variable rate. In the second case, $c$ must
be zero matrix, which the model becomes the one considered in
\cite{13}, that is the extended drop-push model. In section 4, we
take $c$ a non-diagonal matrix and obtain the necessary and
sufficient conditions (constraints) for the matrices $b$ and $c$,
to have a consistent evolution equations. We see that the
reactions are those indicated in eq.(\ref{3}). We find some
classes of solutions of this set of constraints. In section 5, we
investigate the coordinate Bethe ansatz solution for our problem
and obtain the SYB equation, to ensure the solvability of the
model. For $p=2$, we show that for a specific class of parameters,
which corresponds to a special exchange-reaction together with a
special extended drop-push model, the SYB equation has a unique
solution. Finally in section 6, the conditional two-particle
probabilities are calculated for this unique model, and their
large-time behaviour are studied.
\section{The general boundary condition}
Consider a $p$-species system with particles $A_1,A_2,\cdots
,A_p$. The basic quantities that we are interested in are the
probabilities $P_{\alpha _1\cdots\alpha _N}(x_1,\cdots ,x_N;t)$
for finding at time $t$ the particle of type $\alpha _1$ at site
$x_1$, particle of type $\alpha _2$ at site $x_2$, etc.. We take
these
functions to define probabilities only in the physical region $%
x_1<x_2<...<x_N$. The most general master equation for an
asymmetric exclusion process is
 \bea\label{4}
 {\partial\over{\partial t}} P_{\alpha _1\cdots\alpha_N} (x_1,\cdots ,x_N;t)
 &=& P_{\alpha _1\cdots\alpha _N}(x_1-1,\cdots ,x_N;t)
 + \cdots + P_{\alpha _1\cdots\alpha _N}(x_1,\cdots ,x_N-1;t) \cr &&
 -NP_{\alpha _1\cdots\alpha _N}(x_1,\cdots ,x_N;t).
 \eea
This equation describes a collection of $N$ particles drifting to
the right with unit rate. It can be shown that if one decides to
solve the master equation by coordinate Bethe ansatz method, one
may only choose all the diffusion rates equal, which can be scaled
to one \cite{11}. So in eq.(\ref{4}), we take all the diffusion
rates equal to one. The master equation (\ref{4}) is only valid
for
 \be\label{5}
 x_i<x_{i+1}-1,
 \end{equation}
since for $x_i=x_{i+1}-1$, there will be terms with $x_i=x_{i+1}$
in the right-hand side of eq.(\ref{4}), which is out of the
physical region. One can, however, assume that (\ref{4}) is
correct for all the physical region $x_i<x_{i+1}$, and impose
certain boundary conditions for $x_i=x_{i+1}$. These boundary
conditions determine the nature of the interactions between
particles. Now the question is that what are the possible boundary
conditions? To see this, we follow the same argument which has
been given for single-species model in \cite{14}. If one considers
the master equations (\ref{4}) for two-particles probabilities,
finds
 \be\nonumber
  {\partial \over{\partial
 t}}\sum_{x_2}\sum_{x_1<x_2}P_{\a_1\a_2}(x_1,x_2;t)=
 \sum_{x_2}\sum_{x_1<x_2}[ P_{\a_1\a_2}(x_1-1,x_2;t)
 +P_{\a_1\a_2}(x_1,x_2-1;t)-2 P_{\a_1\a_2}(x_1,x_2;t) ]
 \end{equation}
 \bea\label{6}
 &=& \sum_{x_2}\sum_{x_1<x_2}P_{\a_1\a_2}(x_1,x_2;t) -
 \sum_x P_{\a_1\a_2}(x,x+1;t)+ \sum_{x_2}\sum_{x_1<x_2}
 P_{\a_1\a_2}(x_1,x_2;t) \cr &&+
 \sum_xP_{\a_1\a_2}(x,x;t)-2
 \sum_{x_2}\sum_{x_1<x_2} P_{\a_1\a_2}(x_1,x_2;t)\cr
 &=& -\sum_xP_{\a_1\a_2}(x,x+1;t) +\sum_x P_{\a_1\a_2}(x,x;t).
 \eea
Now let us exclude the creation and annihilation processes, in
other words we consider the processes in which the number of
particles is constant in time. Therefore if we sum eq.(\ref{6})
over $\a_1$ and $\a_2$, the left-hand side becomes zero, so the
right-hand side must be also zero. The only possible choice that
leads the right-hand side of eq.(\ref{6}) to zero, and consistent
with more-than-two particle analogue of eq.(\ref{6}) (see
ref.\cite{14}) is taking $P_{\a_1\a_2}(x,x;t)$ as a linear
combination of $P_{\b_1\b_2}(x,x+1;t)$ and
$P_{\b_1\b_2}(x-1,x;t)$'s. Therefore the most general boundary
condition is
 \be\label{7}
 P_{\a_1\a_2}(x,x) =\sum_\b b^{\b_1\b_2}_{\a_1\a_2}
 P_{\b_1\b_2}(x-1,x) +\sum_\b c^{\b_1\b_2}_{\a_1\a_2}
 P_{\b_1\b_2}(x,x+1).
 \end{equation}
$\b$ stands for $(\b_1\b_2)$. These $b$ and $c$ matrices introduce
interactions to particles. Inserting eq.(\ref{7}) in the
right-hand side of eq.(\ref{6}) and sum over $\a_1$ and $\a_2$,
results
 \be\label{8}
 -\sum_x\sum_\a P_{\a_1\a_2}(x,x+1)+
 \sum_x\sum_\b\left(\sum_\a(b+c)^{\b_1\b_2}_{\a_1\a_2}\right)
 P_{\b_1\b_2}(x,x+1)=0.
 \end{equation}
Therefore the sum over the elements of each column of the matrix
$b+c$ must be equal to one,
 \be\label{9}
 \sum_\a(b+c)^{\b_1\b_2}_{\a_1\a_2}=1.
 \end{equation}
Note that for annihilation process, the condition (\ref{9}) need
not be satisfied \cite{14}.

If the matrices $b$ and $c$ are diagonal, then the boundary
condition (\ref{7}) does not induce reactions in which the types
of the particles are changed. So the left-hand side of (\ref{6})
is zero, without summing over $\a_1$ and $\a_2$, and therefore
 \be\label{10}
 -\sum_x P_{\a_1\a_2}(x,x+1)+
 \sum_x(b+c)^{\a_1\a_2}_{\a_1\a_2}
 P_{\a_1\a_2}(x,x+1)=0.
 \end{equation}
So in diagonal case, we have, instead of condition (\ref{9}),
 \be\label{11}
 b=1-c.
 \end{equation}
The simple exclusion processes of \cite{7} is an example of this
model with $p=1$ and $b=0$, the drop-push model with equal rate is
$p=1$ and $c=0$ case, and with non-equal rate, is an example of
one-species case of eqs. (\ref{7}) and (\ref{11}) \cite{8}. The
two-species model of \cite{10}, special $p$-species of \cite{11}
and general $p$-species reactions of \cite{12} are examples of
eqs. (\ref{7}) and (\ref{9}) with $b=0$. Finally the extended
drop-push model of \cite{13} is an example with $c=0$. Now we are
going to consider the situations which have not been studied yet,
that is the multi-species boundary conditions in which both $b$
and $c$ matrices present.
\section{Diagonal c}
As the first step of our study, let us consider the matrix $c$ to
be diagonal. As we will show in the next section, the non-diagonal
elements of $c$ are reaction rates and so they are non-negative,
but the diagonal elements are not reaction rates and are defined
through eq.(\ref{9}), which for diagonal $c$ results
 \be\label{12}
 c^{\b_1\b_2}_{\b_1\b_2}=1-\sum_\a b^{\b_1\b_2}_{\a_1\a_2}.
 \end{equation}
Now consider ${\dot P}_{\a_1\a_2}(x,x+1)$. Using eqs.(\ref{4}),
(\ref{7}) and (\ref{12}), one finds
 \bea\label{13}
 {\dot P}_{\a_1\a_2}(x,x+1)&=&P_{\a_1\a_2}(x-1,x+1)+
 \sum_\b b^{\b_1\b_2}_{\a_1\a_2}P_{\b_1\b_2}(x-1,x)\cr
 &&-(1+\sum_\b b^{\a_1\a_2}_{\b_1\b_2})P_{\a_1\a_2}(x,x+1).
 \eea
The above evolution equation determines the following reactions as
the source and sink of this model
 \bea\label{14}
 A_\alpha\emptyset &\rightarrow &\emptyset A_\alpha \ \ \ \ {\rm with
 \ rate}\ \ 1,\cr
  A_{\b_1} A_{\b_2}\emptyset &\rightarrow &\emptyset A_{\a_1}A_{\a_2} \ \ \ \ {\rm with
  \ rate} \ \ b^{\b_1\b_2}_{\a_1\a_2}.
 \eea
To check the consistency of our description, we next consider the
3-particle probability $P_{\a_1\a_2\a_3}(x-1,x,x+1)$. In this
case, the boundary term $P_{\a_1\a_2\a_3}(x-1,x,x)$ appears in the
right-hand side of master equation. To calculate this term, we
must use the boundary condition (\ref{7}) in two steps. The result
is
 \bea\label{15}
 P_{\a_1\a_2\a_3}(x-1,x,x)&=& \sum_\b b^{\b_2\b_3}_{\a_2\a_3} [\sum_\g b^{\g_1\g_2}_{\a_1\b_2}
 P_{\g_1\g_2\b_3}(x-2,x-1,x)+c^{\a_1\b_2}_{\a_1\b_2}
 P_{\a_1\b_2\b_3}(x-1,x,x)]\cr &&+ c^{\a_2\a_3}_{\a_2\a_3}
 P_{\a_1\a_2\a_3}(x-1,x,x+1).
 \eea
It is seen that the boundary term $P_{\a_1\a_2\a_3}(x-1,x,x)$ is
found as a linear combination of other boundary terms, i.e.
$P_{\a_1\b_2\b_3}(x-1,x,x)$'s. To avoid this problem, which in
general does not lead to a consistent description of the
interactions, we have two choices. One choice is to restrict
ourselves to the cases in which the multiplication factors of the
boundary terms in the right-hand side of eq.(\ref{15}) are zero.
Then
 \be\label{16}
 c^{\a_1\b_2}_{\a_1\b_2}b^{\b_2\b_3}_{\a_2\a_3}=(1-\sum_\g
 b^{\a_1\b_2}_{\g_1\g_2})b^{\b_2\b_3}_{\a_2\a_3}=0.
 \end{equation}
So for non-zero reaction rates, i.e. $b^{\b_2\b_3}_{\a_2\a_3}\neq
0$, we have $c^{\a_1\b_2}_{\a_1\b_2}=0$ (i.e. $\sum_\g
 b^{\a_1\b_2}_{\g_1\g_2}=1$). In this case the model reduced to
one studied in \cite{13}, that is the extended drop-push model.

The second choice is to consider the situations in which the
boundary term in the right-hand side of eq.(\ref{15}) is the same
as one appeared in the left-hand side. In other words, we choose
 \be\label{17}
 c^{\a_1\b_2}_{\a_1\b_2}b^{\b_2\b_3}_{\a_2\a_3}= r_{\a_1\a_2\a_3}
 \delta^{\b_2}_{\a_2}\delta^{\b_3}_{\a_3}.
 \end{equation}
This means that the matrix $b$ must be also diagonal. Therefore we
have
 \bea\label{18}
 b_{\a\b}^{\a\b}:=r_{\a\b} \ \ \ &,& \ \ \
 c_{\a\b}^{\a\b}=1-r_{\a\b}\cr
 P_{\a\b}(x,x)=r_{\a\b}P_{\a\b}(x-1,x)&+& (1-r_{\a\b})
 P_{\a\b}(x,x+1).
 \eea
Considering ${\dot P}_{\a\b}(x,x+1)$, eqs.(\ref{4}) and (\ref{18})
show that the reactions are
 \bea\label{19}
 A_\alpha\emptyset &\rightarrow &\emptyset A_\alpha \ \ \ \ {\rm with
 \ rate}\ \ 1,\cr
  A_{\a} A_{\b}\emptyset &\rightarrow &\emptyset A_{\a}A_{\b} \ \ \ \ {\rm with
  \ rate} \ \ r_{\a\b}.
 \eea
But if one considers ${\dot P}_{\a\b\g}(x,x+1,x+2)$, or other
$n$-adjacent particles probabilities, finds that the master
equation (\ref{4}), boundary condition (\ref{7}), with $b$ and $c$
given in (\ref{18}) can consistently describe the drop-push
reactions only if $r_{\a\b}$ is independent of $\a$ and $\b$. In
this way we arrive at the one-parameter drop-push model,
previously studied in \cite{8}.

So in brief, considering $c$ as a diagonal matrix, does not lead
to any new model.
\section{Non-diagonal c}
Like the previous section, to find the reactions which exist in
this case, we must first consider ${\dot P}_{\a_1\a_2}(x,x+1)$.
Using eqs.(\ref{4}), (\ref{7}) and (\ref{9}), we arrive at
  \bea\label{20}
 {\dot P}_{\a_1\a_2}(x,x+1)&=&P_{\a_1\a_2}(x-1,x+1)+
 \sum_\b b^{\b_1\b_2}_{\a_1\a_2}P_{\b_1\b_2}(x-1,x)\cr
 &&+\sum_\b c_{\a_1\a_2}^{\b_1\b_2}P_{\b_1\b_2}(x,x+1)
 -2P_{\a_1\a_2}(x,x+1)\cr &=& P_{\a_1\a_2}(x-1,x+1)+
 \sum_\b b^{\b_1\b_2}_{\a_1\a_2}P_{\b_1\b_2}(x-1,x)
 +\sum_{\b\neq \a} c_{\a_1\a_2}^{\b_1\b_2}P_{\b_1\b_2}(x,x+1)\cr
 &&-( 1+\sum_\b b_{\b_1\b_2}^{\a_1\a_2}+ \sum_{\b\neq \a}
 c^{\a_1\a_2}_{\b_1\b_2})P_{\a_1\a_2}(x,x+1),
 \eea
in which we use
 \be\label{21}
 c^{\a_1\a_2}_{\a_1\a_2}= 1- \sum_\b b_{\b_1\b_2}^{\a_1\a_2}- \sum_{\b\neq \a}
 c^{\a_1\a_2}_{\b_1\b_2}.
 \end{equation}
The evolution equation (\ref{20}) describes the following
two-particle reactions:
 \bea\label{22}
 A_\alpha\emptyset &\rightarrow &\emptyset A_\alpha \ \ \ \ {\rm with
 \ rate}\ \ 1,\cr
  A_\alpha A_\beta &\rightarrow &A_\gamma A_\delta  \ \ \ \ {\rm with
  \ rate} \ \ c^{\alpha\beta}_{\gamma\delta},\cr
  A_\alpha A_\beta\emptyset &\rightarrow &\emptyset A_\gamma A_\delta  \ \ \ \ {\rm with
  \ rate} \ \ b^{\alpha\beta}_{\gamma\delta}.
  \eea
To find the more-than-two particles reactions, we must consider
${\dot P}_{\a_1\cdots\a_n}(x,x+1,\cdots ,x+n-1)$. In $n=3$, we
first need to know $P_{\a_1\a_2\a_3}(x-1,x,x)$. Using (\ref{7}),
we have
  \bea\label{23}
 P_{\a_1\a_2\a_3}(x-1,x,x)&=& \sum_{\b\g} b^{\b_2\b_3}_{\a_2\a_3} [b^{\g_1\g_2}_{\a_1\b_2}
 P_{\g_1\g_2\b_3}(x-2,x-1,x)+c^{\g_1\g_2}_{\a_1\b_2}
 P_{\g_1\g_2\b_3}(x-1,x,x)]\cr &&+ \sum_\b c^{\b_2\b_3}_{\a_2\a_3}
 P_{\a_1\b_2\b_3}(x-1,x,x+1).
 \eea
Again we consider two cases. In the first case, we take the
matrices $b$ and $c$ such that
  \be\label{24}
 \sum_{\b_2} c^{\g_1\g_2}_{\a_1\b_2}b^{\b_2\b_3}_{\a_2\a_3}= r_{\a_1\a_2\a_3}
 \delta^{\g_1}_{\a_1} \delta^{\g_2}_{\a_2}\delta^{\b_3}_{\a_3}.
 \end{equation}
Then eq.(\ref{23}) results
 \be\label{25}
 P_{\vec
 \a}(x-1,x,x)=\sum_{\b\g}{b_{\a_1\g}^{\b_1\b_2}b_{\a_2\a_3}^{\g\b_3}\over
 {1-r_{\vec\a}}}P_{\vec \b}(x-2,x-1,x)
 +\sum_\b{c_{\a_2\a_3}^{\b_2\b_3}\over
 {1-r_{\vec\a}}}P_{\a_1\b_2\b_3}(x-1,x,x+1),
  \end{equation}
where ${\vec \a}=(\a_1,\a_2,\cdots )$. Using (\ref{25}) in
computing ${\dot P}_{\vec \a}(x-1,x,x+1)$, one can easily see that
the resulting evolution equation gives different rates for same
reaction. For example it gives the rate $c_{\a_1\a_2}^{\b_1\b_2}$
for process $\b_1\b_2\a_3\rightarrow\a_1\a_2\a_3$, which is
consistent with second reaction of eq.(\ref{22}), and rate
$c_{\a_2\a_3}^{\b_2\b_3}/(1-r_{\vec \a})$ for reaction
$\a_1\b_2\b_3\rightarrow\a_1\a_2\a_3$. The only way to obtain the
consistent result is taking $r_{\vec \a}=0$. Therefore we restrict
ourselves to the matrices $b$ and $c$ with property
 \be\label{26}
 \sum_{\b_2} c^{\g_1\g_2}_{\a_1\b_2}b^{\b_2\b_3}_{\a_2\a_3}=0.
 \end{equation}
Note that this constraint is equivalent to
 \be\label{27}
 (1\otimes b)(c\otimes 1)=0,
 \end{equation}
in which $b$ and $c$ are $p^2\times p^2$ matrices satisfying
(\ref{9}), and $1$ denotes the $p\times p$ identity matrix.
Assuming (\ref{26}) and using eqs.(\ref{4}), (\ref{7}) and,
(\ref{23}), ${\dot P}_{\vec \a}(x,x+1,x+2)$ becomes
 \bea\label{27}
 {\dot P}_{\vec \a}(x,x+1,x+2) &=&P_{\vec \a}(x-1,x+1,x+2)
 +\sum_\b b^{\b_1\b_2}_{\a_1\a_2}P_{\b_1\b_2\a_3}(x-1 ,x,x+2) \cr
 &&+\sum_{\b\neq\a} c^{\b_1\b_2}_{\a_1\a_2}P_{\b_1\b_2\a_3}(x, x+1,x+2)
 +\sum_\b b^{\vec\b}_{\vec\a}P_{\vec\b}(x-1 ,x,x+1)\cr &&
 +\sum_{\b\neq\a} c^{\b_2\b_3}_{\a_2\a_3}P_{\a_1\b_2\b_3}(x ,x+1,x+2)
-(1+\sum_\b b^{\a_1\a_2}_{\b_1\b_2} +\sum_{\b\neq\a}
c^{\a_1\a_2}_{\b_1\b_2}\cr &&
 +\sum_\b b^{\a_2\a_3}_{\b_2\b_3} +\sum_{\b\neq\a}
 c^{\a_2\a_3}_{\b_2\b_3}) P_{\vec \a}(x,x+1,x+2),
 \eea
in which we use eq.(\ref{21}) for $c^{\a_1\a_2}_{\a_1\a_2}$ and
$c^{\a_2\a_3}_{\a_2\a_3}$. In above equation,
$b^{\vec\b}_{\vec\a}$ stands for
 \be\label{28}
 b^{\vec\b}_{\vec\a}=\sum_\g b^{\b_1\b_2}_{\a_1\g}
 b^{\g\b_3}_{\a_2\a_3}.
 \end{equation}
The source terms of eq.(\ref{27}), besides the reactions
(\ref{22}), are the following 3-particle drop-push reaction
 \be\label{29}
 A_{\b_1}A_{\beta_2}A_{\b_3}\emptyset \rightarrow\emptyset A_{\a_1}A_{\a_2}A_{\a_3} \ \ \ \ {\rm with
  \ rate} \ \ b^{\vec\b}_{\vec\a}.
 \end{equation}
The sink terms are also consistent with reactions (\ref{22}) and
(\ref{29}), provided
 \be\label{30}
 \sum_\b b^{\vec\a}_{\vec\b}=\sum_{\b\g} b^{\a_1\a_2}_{\b_1\g}
 b^{\g\a_3}_{\b_2\b_3}=\sum_\b b^{\a_1\a_2}_{\b_1\b_2}.
 \end{equation}
This is the last constraints that must be satisfied by the
elements of matrix $b$. It can be shown that the more-than-three
adjacent particles probabilities are consistent with following
reactions

  \bea\label{31}
 A_\alpha\emptyset &\rightarrow &\emptyset A_\alpha \ \ \ \ {\rm with
 \ rate}\ \ 1,\cr
  A_\alpha A_\beta &\rightarrow &A_\gamma A_\delta  \ \ \ \ {\rm with
  \ rate} \ \ c^{\alpha\beta}_{\gamma\delta},\cr
  A_{\a_0}\cdots A_{\a_n}\emptyset &\rightarrow &\emptyset A_{\g_0}\cdots A_{\g_n}  \ \ \ \ {\rm with
  \ rate} \ \ (b_{n-1,n}\cdots
  b_{0,1})^{\a_0\cdots\a_n}_{\g_0\cdots\g_n},
  \eea
if the constraints (\ref{9}), (\ref{26}) and, (\ref{30}) are
satisfied. In eq.(\ref{31}), we use the following definition
 \be\label{32}
 b_{k,k+1}=1\otimes\cdots\otimes
 1\otimes\underbrace{b}_{k,k+1}\otimes 1 \otimes\cdots \otimes 1.
 \end{equation}
The important point is that we need not any further constraint.
For example in describing the source and sink terms of the
evolution equation of $P_{\vec\a}(x,x+1,x+2,x+3)$, we encounter
the constraint
 \be\label{c}
  \sum_{\b\g\theta}
 b^{\a_1\a_2}_{\b_1\g} b^{\g\a_3}_{\b_2\theta}
 b^{\theta\a_4}_{\b_3\b_4}=\sum_\b b^{\a_1\a_2}_{\b_1\b_2}.
 \end{equation}
Now as ( see eq.(\ref{30}))
 \be
 \sum_{\b\theta} b^{\g\a_3}_{\b_2\theta}
 b^{\theta\a_4}_{\b_3\b_4}=\sum_\b b^{\g\a_3}_{\b_2\b_3},
  \end{equation}
constraint (\ref{c}) reduces to $\sum_{(\b\g)}
b^{\a_1\a_2}_{\b_1\g} b^{\g\a_3}_{\b_2\b_3} =\sum_\b
b^{\a_1\a_2}_{\b_1\b_2}$, which is nothing but eq.(\ref{30}). This
completes our investigation and the final processes are those
indicated in (\ref{31}). Of course, the question of solvability is
open yet.

As the elements of matrix $b$ and the non-diagonal elements of
matrix $c$ are reaction rates, they must be non-negative, and this
point restricts the allowed solutions of eqs.(\ref{9}), (\ref{26})
and (\ref{30}). Let us focus on two class of solutions. To be
specific, we consider $p=2$ case, but the arguments can be easily
applied to arbitrary $p$. In $p=2$, we may denote the two-particle
states $(\a_1,\a_2)$ as following:
 \be\label{33}
 |1>=(1,1) \ , \ |2>=(1,2) \ , \ |3>=(2,1) \ , \ |4>=(2,2).
 \end{equation}
{\bf solution of class 1:} If $b^{\b_2\b_3}_{\a_2\a_3}$ in
eq.(\ref{26}) is independent of $\b_2$, that is
 \be\label{34}
 b^{1\b_3}_{\a_2\a_3}=b^{2\b_3}_{\a_2\a_3}=\cdots
 =b^{p\b_3}_{\a_2\a_3},
 \end{equation}
then eq.(\ref{26}) reduces to
 \be\label{35}
 \sum_\b c^{\g_1\g_2}_{\a_1\b}=0.
 \end{equation}
For $p=2$, eq.(\ref{34}) gives $b_i^{\ 1}=b_i^{\ 3}$ ( for
$\b_3=1$) and $b_i^{\ 2}=b_i^{\ 4}$ ( for $\b_3=2$), and
eq.(\ref{35}) for $\a_1=1,2$ specifies the elements of $c$ as
following
 \begin{equation}\label{36}
 c=\left(
 \begin{array}{cccc}
 -c_2^{\ 1} &c_1^{\ 2} & 0 & 0 \\
 c_2^{\ 1} &-c_1^{\ 2} &0 & 0 \\
 0 & 0 &-c_4^{\ 3} & c_3^{\ 4}\\
 0 & 0 & c_4^{\ 3} &-c_3^{\ 4}
 \end{array}
 \right) .
 \end{equation}
It can be easily shown that if we take the sum of elements of each
column of $b$ equal to one, then the constraints (\ref{9}) and
(\ref{30}) are also satisfied. The resulting model is a
ten-parameter reaction-diffusion model.\\
{\bf solution of class 2:} If we take, for each $\b$, one or both
of $c^{\g_1\g_2}_{\a_1\b}$ and $b^{\b\g_3}_{\a_2\a_3}$ equal to
zero, then eq.(\ref{26}) is satisfied. In $p=2$, we can take
$c^{\g_1\g_2}_{\a_12}=0$ and $b^{1\g_3}_{\a_2\a_3}=0$, which means
that taking zero the second and fourth rows of $c$ and first and
second columns of $b$. Then solving (\ref{30}), results $b_2^{\
3}+b_4^{\ 3}=b_2^{\ 4}+b_4^{\ 4}=1$ and $\sum_ib_i^{\
3}=\sum_ib_i^{\ 4}$. Finally considering (\ref{9}) and noting the
non-negativity of reaction rates, gives
 \be\label{37}
 b=\left(
 \begin{array}{cccc}
  0&0& 0 & 0 \\
  0&0&b_2^{\ 3} &b_2^{\ 4} \\
 0 & 0 &0&0\\
 0 & 0 & 1-b_2^{\ 3} &1-b_2^{\ 4}
 \end{array}
 \right) \ , \
 c=\left(
 \begin{array}{cccc}
 1-c_3^{\ 1} &c_1^{\ 2} &c_1^{\ 3}  & 0 \\
 0 &0 &0 & 0 \\
 c_3^{\ 1} &1-c_1^{\ 2} &- c_1^{\ 3}&0\\
 0 & 0 & 0&0
 \end{array}
 \right),
 \end{equation}
with condition
 \be\label{38}
 (b_2^{\ 3},b_2^{\ 4},c_1^{\ 2})\leq 1.
 \end{equation}
In $p=2$, one can show that these two classes are the complete set
of solutions of the constraint equations. Now we must seek the
Bethe ansatz solution for the allowed set of parameters.
\section{The Bethe ansatz solution}
To solve the master equation (\ref{4}) with boundary condition
(\ref{7}), we consider the following ansatz
 \begin{equation}\label{39}
P_{\alpha_1,\cdots ,\alpha _N}({\mathbf x};t)=e^{-E_Nt}\psi
_{\alpha_1,\cdots,\alpha _N}({\mathbf x}),
\end{equation}
with
 \begin{equation}\label{40}
\Psi({\mathbf x})=\sum_\sigma {\mathbf{A}}_\sigma e^{i\sigma
({\mathbf{p}}).{\mathbf{x}}}.
\end{equation}
$\Psi$ is a tensor of rank $N$ with components $\psi
_{\alpha_1,\cdots,\alpha _N}({\mathbf x})$, and the summation runs
over the elements of the permutation group of $N$ objects
\cite{n10}\cite{n11}. Inserting (\ref{39}) in eqs.(\ref{4}) and
(\ref{7}), results, respectively,
 \be\label{41}
 E_N=\sum_{k=1}^N(1-e^{-ip_k}),
 \end{equation}
and
 \bea\label{42}
  \Psi(\cdots ,x_k=x,x_{k+1}=x,\cdots&)&=b_{k,k+1}
  \Psi(\cdots ,x_k=x-1,x_{k+1}=x,\cdots ) \cr &+&c_{k,k+1}
  \Psi(\cdots ,x_k=x,x_{k+1}=x+1,\cdots ).
 \eea
$c_{k,k+1}$ defined like $b_{k,k+1}$ in eq.(\ref{32}). The
coefficients ${\mathbf{A}}_\sigma$ can be determined by inserting
eq.(\ref{40}) in (\ref{42}), which gives
 \be\label{44}
 [1-e^{-i\sigma (p_k)}b_{k,k+1}- e^{i\sigma (p_{k+1})}c_{k,k+1} ]
 {\mathbf A}_\sigma +
 [1-e^{-i\sigma (p_{k+1})}b_{k,k+1}- e^{i\sigma (p_{k})}c_{k,k+1} ]
 {\mathbf A}_{\sigma\sigma_k}=0.
 \end{equation}
$\sigma_k$ is an element of permutation group which only
interchanges $p_k$ and $p_{k+1}$:
 \be\label{i}
 \s_k:(p_1,\cdots ,p_k,p_{k+1},\cdots ,p_N)\rightarrow
 (p_1,\cdots ,p_{k+1},p_{k},\cdots ,p_N).
 \end{equation}
Using (\ref{44}), ${\mathbf A}_{\sigma\sigma_k}$ is obtained from
${\mathbf A}_{\sigma}$ as following
 \be\label{45}
 {\mathbf A}_{\sigma\sigma_k}=S_{k,k+1}(\s (p_k),\s (p_{k+1}))
 {\mathbf A}_{\sigma},
 \end{equation}
where
 \be\label{46}
 S_{k,k+1}(z_1,z_2)=1\otimes\cdots\otimes
 1\otimes\underbrace{S(z_1,z_2)}_{k,k+1}\otimes 1 \otimes\cdots \otimes 1.
 \end{equation}
$S(z_1,z_2)$ is the following $p^2\times p^2$ matrix
 \be\label{47}
 S(z_1,z_2)=-(1-z_2^{-1}b-z_1c)^{-1}(1-z_1^{-1}b-z_2c),
 \end{equation}
and $z_k=e^{ip_k}$.

Until now, all the $b$ and $c$ matrices are acceptable. But we
must note that the generators of the permutation group satisfy
$\s_k\s_{k+1}\s_k=\s_{k+1}\s_k\s_{k+1}$, so one also needs
 \be\label{48}
 {\mathbf A}_{\s_k\s_{k+1}\s_k}={\mathbf A}_{\s_{k+1}\s_k\s_{k+1}}.
 \end{equation}
In terms of $S$-matrices, eq.(\ref{48}) becomes
 \be\label{49}
 S_{12}(z_2,z_3)S_{23}(z_1,z_3)S_{12}(z_1,z_2)=
 S_{23}(z_1,z_2)S_{12}(z_1,z_3)S_{23}(z_2,z_3).
 \end{equation}
Writing $S$-matrix as the product of the permutation matrix $\Pi$
and a $R$ matrix:
 \be\label{50}
 S_{k,k+1}=:\Pi_{k,k+1}R_{k,k+1},
 \end{equation}
eq.(\ref{49}) is transformed to
 \be\label{51}
 R_{23}(z_2,z_3)R_{13}(z_1,z_3)R_{12}(z_1,z_2)=
 R_{12}(z_1,z_2)R_{13}(z_1,z_3)R_{23}(z_2,z_3).
 \end{equation}
This is the spectral Yang-Baxter equation \cite{n12}-\cite{n14}.

Now the matrices $b$ and $c$ must be such that if one computes the
two-particle $S$-matrix (\ref{47}) by them, it satisfies the
spectral Yang-Baxter  equation (\ref{51}), or equivalently
eq.(\ref{49}). These equations are complicated and determining the
elements of $b$ and $c$, in such a way that eq.(\ref{49}) is
satisfied, is not easy.

For $b=0$, the $S$-matrix (\ref{47}) becomes a binomial of degree
one with respect to $z_2$ and therefore (\ref{49}) becomes a
quadratic expression with respect to $z_3$. Using this, one can
transform the spectral Yang-Baxter equation (\ref{49}) to a
non-spectral matrix equation for matrix $c$, which its study is
much easier than eq.(\ref{49}) \cite{12}. The same is true for
$c=0$, since in this case eq.(\ref{47}) is linear in
$z_1^{-1}=e^{-ip_1}$ \cite{13}. When $b$ and $c$ are both
different from zero, this procedure does not work because of the
presence of $z_2$ and $z_2^{-1}$ ( or equally $z_1$ and
$z_1^{-1}$) in both terms of (\ref{47}).

For matrices $b$ and $c$ which are given in eq.(\ref{37}), if one
computes the corresponding $S$-matrix and writes eq.(\ref{49}) as
RHS - LHS = 0, then one has 64 equations that must be solved for
five variables $b_2^{\ 3},b_2^{\ 4},c_1^{\ 2},c_1^{\ 3}$ and
$c_3^{\ 1}$. The solutions must be momentum-independent and
non-negative. By obtaining the set of solutions of this system of
equations by standard mathematical softwares, it is seen that all
of the solutions are momentum-dependent except one. This only
acceptable solution is: $b_2^{\ 3}=1$, and other four parameters
are zero. Therefore the solvable two-species reaction-diffusion
model is defined through the following $b$ and $c$ matrices:
 \be\label{52}
 b=\left(
 \begin{array}{cccc}
  0&0& 0 & 0 \\
  0&0&1&0 \\
 0 & 0 &0&0\\
 0 & 0 & 0&1
 \end{array}
 \right) \ , \
 c=\left(
 \begin{array}{cccc}
 1&0&0& 0 \\
 0 &0 &0 & 0 \\
 0&1&0&0\\
 0 & 0 & 0&0
 \end{array}
 \right).
 \end{equation}
These matrices introduce the following reactions, with $A\equiv
A_1$ and $B\equiv A_2$,
 \bea\label{53}
 A\emptyset &\rightarrow & \emptyset A \cr
 B\emptyset &\rightarrow & \emptyset B \cr
 AB &\rightarrow & BA \cr
 BA\emptyset &\rightarrow & \emptyset AB \cr
 BB\emptyset &\rightarrow & \emptyset BB \cr
 BAA\emptyset &\rightarrow & \emptyset AAB \cr
 BAB\emptyset &\rightarrow & \emptyset ABB \cr
 BBA\emptyset &\rightarrow & \emptyset BAB \cr
 BBB\emptyset &\rightarrow & \emptyset BBB \cr
 &\vdots&
 \eea
The dots indicates the more-than-three particle drop-push
reactions which are specified by eqs.(\ref{31}) and (\ref{32}),
and all reactions occur with rate one.
\section{Two-particle conditional probabilities for reactions
(\ref{53})} The conditional probability $P({\vec \a},{\mathbf x};
t|{\vec \b},{\mathbf y};0)$ is the probability of finding
particles $\a_1,\a_2,\cdots$ at time $t$ at sites
$x_1,x_2,\cdots$, respectively, if at $t=0$ we have particles
$\b_1,\b_2,\cdots$ at sites $y_1,y_2,\cdots$, respectively. In
two-particle sector, it is
 \begin{equation}\label{54}
  \left(
 \begin{array}{c}
 P_{AA} \\ P_{AB} \\ P_{BA} \\ P_{BB}
 \end{array}
 \right) ({\mathbf{x}};t|{\vec \beta},{\mathbf{y}};0)=\frac
 1{(2\pi )^2}\int e^{-E_2t}e^{-i\mathbf{p.y}}
 \Psi(x_1,x_2)dp_1dp_2,
 \end{equation}
where by eqs.(\ref{40}), (\ref{45}), (\ref{47}) and, (\ref{52}),
$\Psi (x_1,x_2)$ is:
 \be\label{55}
 \Psi (x_1,x_2)=\left( \begin{array}{c}
 a \\ b \\ c \\ d
 \end{array}
 \right) e^{i{\mathbf {p.x}}}+S_{12}(z_1,z_2)\left(
 \begin{array}{c} a \\ b \\ c \\ d
 \end{array}
 \right) e^{i{\widetilde {\mathbf p}}.{\mathbf x}} ,
 \end{equation}
in which ${\mathbf p}=(p_1,p_2)$, ${\widetilde {\mathbf
p}}=(p_2,p_1)$, and $\left(  \begin{array}{c} a \\ b \\ c \\ d
\end{array} \right)$ stands for ${\mathbf A}_{\s =1}$ and
specified by the initial condition. $z_1=e^{ip_1}$,
$z_2=e^{ip_2}$, and $S_{12}(z_1,z_2)$ is:
 \be\label{56}
 S_{12}(z_1,z_2)=\left(
 \begin{array}{cccc}
 {z_2-1 \over {1-z_1}}&0&0& 0 \\
 0 &0 &{1\over z_1} & 0 \\
 0&z_2&0&0\\
 0 & 0 & 0&{z_2(1-z_1)\over z_1(z_2-1)}
 \end{array}
 \right).
 \end{equation}
After some calculations, one finds ( see for example
refs.\cite{10} and \cite{11} for more details):
 \bea\label{58}
 P(A,A,{\mathbf x};t|A,A,{\mathbf y};0)&=&F_1(t)+F_2(t),\cr
 P(A,B,{\mathbf x};t|A,B,{\mathbf y};0)&=&F_1(t),\cr
  P(B,A,{\mathbf x};t|A,B,{\mathbf y};0)&=&F_4(t),\cr
 P(B,A,{\mathbf x};t|B,A,{\mathbf y};0)&=&F_1(t),\cr
 P(A,B,{\mathbf x};t|B,A,{\mathbf y};0)&=&F_3(t),\cr
 P(B,B,{\mathbf x};t|B,B,{\mathbf y};0)&=&F_1(t)+F_5(t),
 \eea
and all other probabilities are zero. $F_i(t)$'s are:
  \bea\label{59}
  F_1(t) & = & e^{-2t}\frac{t^{x_1-y_1}}{(x_1-y_1)!}\frac{t^{x_2-y_2}}
  {(x_2-y_2)!}, \cr
 F_2(t) & = & e^{-2t}\left\{ \frac{t^{x_1-y_2+1}}{(x_1-y_2+1)!}-\frac
 {t^{x_1-y_2}}{(x_1-y_2)!}\right\} \sum\limits_{k=0}^\infty
 \frac{t^{x_2-y_1+k}}{(x_2-y_1+k)!},\cr
 F_3(t) & = & e^{-2t}\frac{t^{x_2-y_1-1}}{(x_2-y_1-1)!}\frac{t^{x_1-y_2}}
  {(x_1-y_2)!}, \cr
 F_4(t) & = & e^{-2t}\frac{t^{x_2-y_1}}{(x_2-y_1)!}\frac{t^{x_1-y_2+1}}
  {(x_1-y_2+1)!}, \cr
 F_5(t) & = & e^{-2t}\left\{ \frac{t^{x_2-y_1}}{(x_2-y_1)!}-\frac
 {t^{x_2-y_1-1}}{(x_2-y_1-1)!}\right\} \sum\limits_{k=1}^\infty
 \frac{t^{x_1-y_2+k}}{(x_1-y_2+k)!}.
 \eea
Note that beginning with initial state $(A,B)$, the system can go
to state $(B,A)$ by third reaction of (\ref{53}) and again goes
back to $(A,B)$ by fourth reaction, etc.. As the rates of
reactions are equal, we expect that at large time, the
probabilities of finding $(A,B)$ and $(B,A)$ are equal, if they
are summed over all accessible sites. In fact, if one calculates
the difference of these two probabilities, finds:
 \bea\label{60}
 D(t)&=:&\sum_{x_2=y_2}^\infty\sum_{x_1=y_1}^{x_2-1}\left[ P(A,B,{\mathbf x};t|A,B,{\mathbf y};0)
 - P(B,A,{\mathbf x};t|A,B,{\mathbf y};0) \right] \cr &=&
 e^{-2t}\left[ 2\sum_{m=1}^{y_2-y_1-1} I_m(2t) +I_0(2t)\right],
 \eea
in which $I_n(x)$ is the $n$-th order Bessel function of the first
kind. Now at $x\rightarrow \infty$, we have
 \be\label{61}
 I_n(x)\rightarrow {e^x\over {\sqrt{2\pi x}}},
 \end{equation}
so
 \be\label{62}
 \lim_{t\rightarrow\infty}D(t)={M\over {\sqrt{4\pi t}}}
 \rightarrow 0,
 \end{equation}
where $M$ is the number of $I_n(2t)$s in eq.(\ref{60}). The final
result confirms our expectation. The same result can be proved
when the initial state is $(B,A)$.

{\bf Acknowledgement:} I would like to thank N. Ahmadi for useful
discussions.


\end{document}